\documentclass[twocolumn]{aastex61}
\usepackage{amsmath}
\usepackage{rotating}
\usepackage{color}
\usepackage{comment}

\accepted{October 10, 2018}
\submitjournal{ApJ}

\shorttitle{NGC~3256 dense gas outflow + Overleaf template article}
\shortauthors{Michiyama et al.}

\begin{document}
\title{ALMA OBSERVATIONS OF HCN AND HCO$^+$ OUTFLOWS IN THE MERGING GALAXY NGC~3256}

\correspondingauthor{Tomonari Michiyama}
\email{t.michiyama@nao.ac.jp}

\author{Tomonari Michiyama}
\affiliation{Department of Astronomical Science, SOKENDAI (The Graduate University of Advanced Studies), 2-21-1 Osawa, Mitaka, Tokyo 181-8588}
\affiliation{National Astronomical Observatory of Japan, National Institutes of Natural Sciences, 2-21-1 Osawa, Mitaka, Tokyo, 181-8588}

\author{Daisuke Iono}
\affiliation{Department of Astronomical Science, SOKENDAI (The Graduate University of Advanced Studies), 2-21-1 Osawa, Mitaka, Tokyo 181-8588}
\affiliation{National Astronomical Observatory of Japan, National Institutes of Natural Sciences, 2-21-1 Osawa, Mitaka, Tokyo, 181-8588}

\author{Kazimierz Sliwa}
\affiliation{Max-Planck-Institut f\"ur Astronomie, K\"onigstuhl 17, 69117 Heidelberg, Germany}

\author{Alberto Bolatto}
\affiliation{Department of Astronomy and Laboratory for Millimeter-Wave Astronomy, University of Maryland, College Park, MD 20742, USA}

\author{Kouichiro Nakanishi}
\affiliation{Department of Astronomical Science, SOKENDAI (The Graduate University of Advanced Studies), 2-21-1 Osawa, Mitaka, Tokyo 181-8588}
\affiliation{National Astronomical Observatory of Japan, National Institutes of Natural Sciences, 2-21-1 Osawa, Mitaka, Tokyo, 181-8588}

\author{Junko Ueda}
\affiliation{National Astronomical Observatory of Japan, National Institutes of Natural Sciences, 2-21-1 Osawa, Mitaka, Tokyo, 181-8588}
\affiliation{Harvard-Smithsonian Center for Astrophysics, 60 Garden Street, Cambridge, MA 02138, USA}

\author{Toshiki Saito}
\affiliation{Max-Planck-Institut f\"ur Astronomie, K\"onigstuhl 17, 69117 Heidelberg, Germany}

\author{Misaki Ando}
\affiliation{Department of Astronomical Science, SOKENDAI (The Graduate University of Advanced Studies), 2-21-1 Osawa, Mitaka, Tokyo 181-8588}
\affiliation{National Astronomical Observatory of Japan, National Institutes of Natural Sciences, 2-21-1 Osawa, Mitaka, Tokyo, 181-8588}

\author{Takuji Yamashita}
\affiliation{Research Center for Space and Cosmic Evolution, Ehime University, 2-5 Bunkyo-cho, Matsuyama, Ehime 790-8577, Japan}

\author{Min Yun}
\affiliation{Department of Astronomy, University of Massachusetts, Amherst, MA 01003, USA}



\begin{abstract}
We report $\sim 2 \arcsec$ resolution Atacama Large Millimeter/submillimeter Array observations of the HCN~(1--0), HCO$^{+}$~(1--0), CO~(1--0), CO~(2--1), and CO~(3--2) lines towards the nearby merging double-nucleus galaxy NGC~3256.
We find that the high density gas outflow traced in HCN~(1--0) and HCO$^{+}$~(1--0) emission is co-located with the diffuse molecular outflow emanating from the southern nucleus, where a low-luminosity active galactic nucleus (AGN) is believed to be the dominant source of the far-infrared luminosity.
On the other hand, the same lines were undetected in the outflow region associated with the northern nucleus, whose primary heating source is likely related to starburst activity without obvious signs of AGN.
Both HCO$^{+}$~(1--0)/CO~(1--0) line ratio (i.e. dense gas fraction) and the CO~(3--2)/CO~(1--0) line ratio are larger in the southern outflow (0.20$\pm$0.04 and 1.3$\pm$0.2, respectively) than in the southern nucleus (0.08$\pm$0.01, 0.7$\pm$0.1, respectively).
By investigating these line ratios for each velocity component in the southern outflow, we find that the dense gas fraction increases and the CO~(3--2)/CO~(1--0) line ratio decreases towards the largest velocity offset.
This suggests the existence of a two-phase (diffuse and clumpy) outflow.
One possible scenario to produce such a two-phase outflow is an interaction between the jet and the interstellar medium, which possibly triggers shocks and/or star formation associated with the outflow.
\end{abstract}




\section{Introduction} \label{sec:intro}
\subsection{Galactic-scale molecular gas outflow}
Active galactic nucleus (AGN) and supernovae feedback both play an important role in the galaxy evolution scenario \citep[e.g.,][]{Springel+05}.
The feedback occurs in the form of galactic scale outflows which can remove the material for future star formation by expelling gas from the nuclear region and/or heating the interstellar gas to high temperatures in the halo (``negative feedback'').
In contrast, recent optical spectroscopic observations have found the evidence of star formation in the outflow \citep{Maiolino+17}, suggesting that outflows can not only regulate star formation in the disk but simultaneously enhance the formation of new massive stars (``positive feedback'').
Numerical simulations suggest the formation of dense molecular gas clumps in the outflow by the interactions between the jet and the impinging interstellar medium (ISM) \citep[e.g.,][]{Zubovas+14, Costa+15, Ferrara+16, Zubovas+17}, leading to the formation of new massive stars \citep[e.g.,][]{Dugan+14, Wagner+16}. 
While outflowing molecular clouds may be destroyed before they reach high velocities, it is shown that molecular clouds are continuously formed inside the outflow \citep{Richings+18}.
Therefore, in both positive and negative feedback scenarios, the molecular outflow plays an important role in galaxy evolution.

Large collections of both supernova explosions and massive stellar winds as well as radiation and jets emanating from AGNs can be the sources of galactic scale molecular outflows \citep[e.g.,][]{Veilleux+05,Fabian+09,Faucher-Giguere+12,Zubovas+12}.
Since both starbursts and AGNs can be triggered during mergers (e.g., \citealt{Hopkins+06}; \citealt{Nara+08}; \citealt{Debuhr+12}; \citealt{Hayward+14}; \citealt{Michiyama+16}), nearby merging galaxies are ideal laboratories to observationally investigate these phenomena \citep[e.g.,][]{Feruglio+10, Cicone+14, Fiore+17, Pereira-Santaella+18}.
Molecular outflows have been detected in several merging galaxies, such as
Mrk~231 \citep{Feruglio+10, Cicone+12, Gonzalez-Alfonso+14, Aalto+15, Feruglio+15, Lindberg+16},
NGC~6240 \citep{Feruglio+13a, Feruglio+13b, Saito+18, Cicone+18},
Arp~220 \citep{Sakamoto+09,Barcos-Munoz+18}, and 
NGC~1614 \citep{Garc+15,Saito+16}.

Up to the present, most of the molecular outflow studies have focused on the CO emission. 
While CO can be an excellent tracer of the diffuse and extended gas in the outflow, it does not trace the densest parts of the molecular clouds where star formation occurs.
A multi-line observation is necessary to investigate the densest parts and to derive the physical properties of the outflow \citep[e.g.,][]{Aalto+12,Aalto+15}.
HCN and HCO$^+$ are often used as dense gas tracers since their critical densities are roughly three orders of magnitude higher than CO for a given $J$.
However, the flux of HCN and HCO$^+$ are generally more than ten times weaker than CO \citep[e.g.,][]{Papa+07, Walter+17}.
The Atacama Large Millimeter/submillimeter Array (ALMA) provides high enough sensitivity to map the extreme velocity components of faint lines.

\subsection{A case study: NGC~3256}
NGC~3256 is a merging luminous infrared galaxy (LIRG) with a far infrared (FIR) luminosity of $L_{\rm FIR} = 10^{11.43}~L_{\sun}$, and the luminosity distance of $D_{\rm L}$ = 35~Mpc \citep{Sanders+03}.
This galaxy has two nuclei separated by 5$\arcsec$($\sim$ 850 pc).
NGC~3256 is closer than the well studied merging ULIRGs; Arp~220 ($D_{\rm L}$ = 70 Mpc) and NGC~6240 ($D_{\rm L}$ = 100 Mpc).
\citet{Sakamoto+14} found extreme velocity components of CO~(1--0) and CO~(3--2) around the two nuclei and they interpret the components as outflows of cold gas from both the northern and southern nuclei. The characteristics of each outflow are summarized as follows;

\noindent
{\bf [Northern outflow]} \\
\citet{Ohyama+15} argue that the northern galaxy (NGC~3256N) is a starburst galaxy.
\citet{Sakamoto+14} suggest the presence of an uncollimated bipolar outflow from the northern nucleus with the maximum outflow velocity of $\sim750$~km~s$^{-1}$ and outflow mass of $\sim6\times10^7~M_\odot$ (hereafter ``northern outflow").
They conclude that the northern outflow can be driven by the starburst around the northern nucleus.

\noindent
{\bf [Southern outflow]} \\
Unlike the northern nucleus, \citet{Ohyama+15} show that the IRAC color and silicate absorption feature are AGN-like at the nucleus of southern galaxy (NGC~3256S). 
X-ray observations show absorption-corrected AGN luminosity in the 2-10 keV band of $(1.2-2.9)\times$10$^{40}$~erg~s$^{-1}$ \citep{Ohyama+15,Lehmer+15}, which is in the range for low-luminosity AGN.
\citet{Sakamoto+14} suggest the presence of a collimated bipolar CO outflow from the southern nucleus with the maximum outflow velocity of $\sim2600$~km~s$^{-1}$ and outflow mass of $\sim~2.5\times10^7~M_\odot$ (hereafter ``southern outflow") associated with the radio jet detected by the Very Large Array \citep{Neff+03}, which is possibly triggered by a weak or recently dimmed AGN in the southern nucleus.
A broad near infrared (near-IR) H$_2$ emission line is detected from the southern outflow \citep{Emonts+14}, suggesting that the outflowing gas is heated to $\sim 2000$ K by shocks or X-rays.
This is possibly due to the interaction between the southern jet and the ISM.
In addition, the optical IFU observations show that the shock is enhanced in NGC~3256 \citep{Rich+11}, and ALMA line survey observation suggests the presence of shock enhancement by an outflow \citep{Harada+18}.

NGC~3256 is an ideal laboratory to compare the properties of two different types of outflows, AGN driven and starburst driven. 
In order to quantify the physical properties of the outflowing gas in NGC~3256, observations of the outflow in different molecules and transitions are necessary.
The main motivation of this work is to identify the dense gas in the molecular gas outflows in NGC~3256, and to investigate how the outflows impact the surrounding ISM.
We present the ALMA Cycle 3 observations of HCN~(1-0), HCO$^{+}$(1-0), and CO~(2-1) emission towards NGC~3256.
We compare the intensity and spatial distribution of dense gas outflows with the previously detected CO outflows.
Furthermore, we estimate the physical properties of outflows under the assumption of non local thermodynamic equilibrium (non-LTE) by using Bayesian likelihood statistical analysis.

The ALMA imaging results are presented in Section \ref{ALMA-sec}.
We identify outflows in Section \ref{ID-OF-sec}.
Section \ref{RADEX-sec} presents the $\tt{RADEX}$ modeling and Bayesian likelihood analysis.
Properties of the northern and southern outflows are discussed in Section \ref{Discussion-sec}.
We summarize our results in Section \ref{Summary-sec}.
Here, the line ratios (in unit of brightness temperature) are defined by the symbol ``$R$" as follows;
$R_{21/10}$~=~CO~(2--1)~/~CO~(1--0),
$R_{32/10}$~=~CO~(3--2)~/~CO~(1--0),
$R_{\rm HCN/CO}$~=~HCN~(1--0)~/~CO~(1--0),
$R_{\rm HCO^+/CO}$~=~HCO$^+$~(1--0)~/~CO~(1--0), and 
$R_{\rm HCN/HCO^+}$~=~HCN~(1--0)~/~HCO$^+$~(1--0).
We adopt $H_0$~=~73~km~s$^{-1}$ Mpc$^{-1}$, $\Omega_M$ = 0.27, and $\Omega_\Lambda$ = 0.73 for all of the analysis throughout this paper.

\section{ALMA data}\label{ALMA-sec}
\begin{figure*}
\begin{center}
\includegraphics[width=17.5cm]{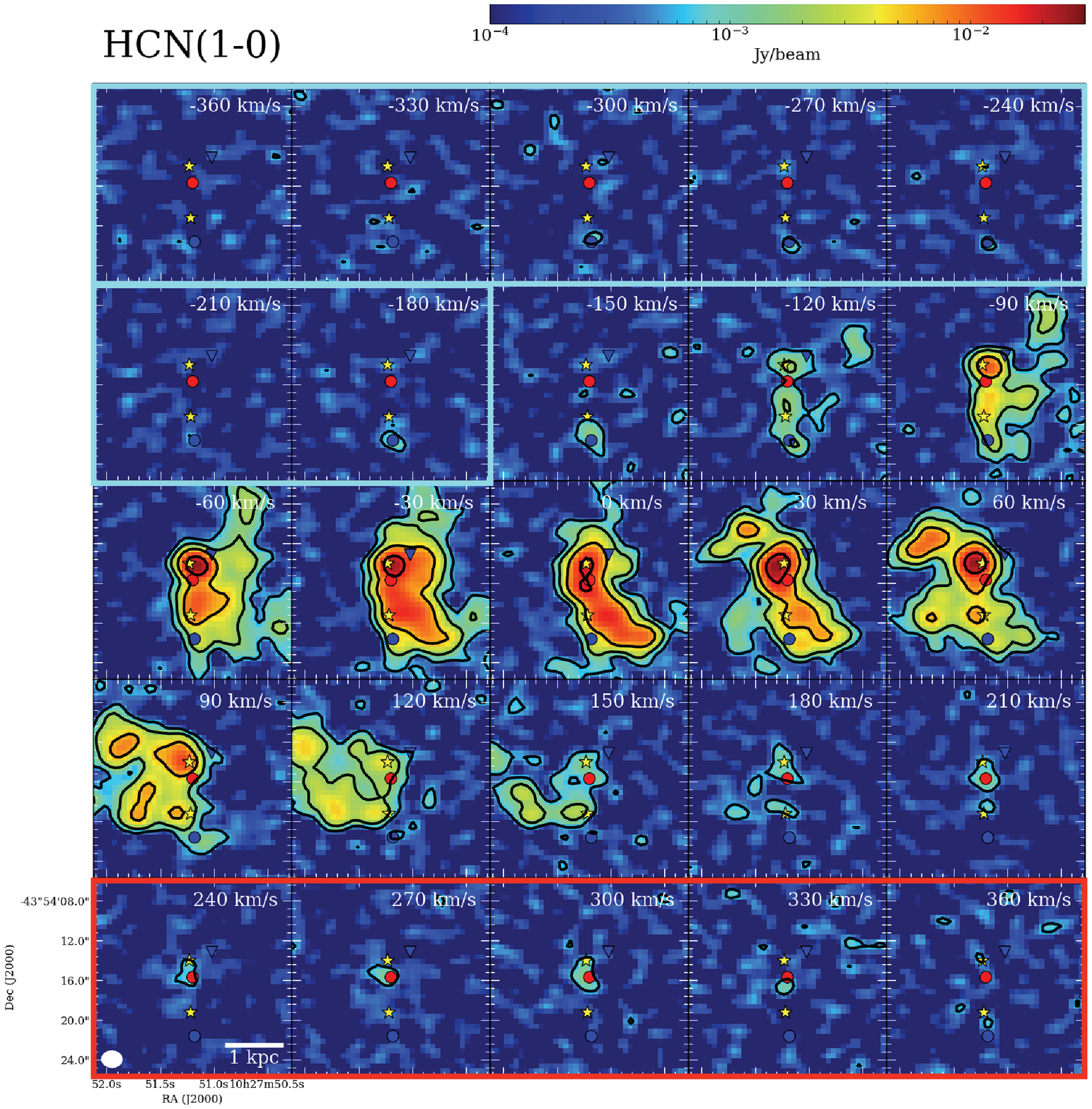}
\caption{
HCN~(1--0) channel map.
The size of each map is 20~\arcsec (3.4~kpc) square centered on
$\rm{RA.}=10^{\rm{h}} 27^{\rm{m}} 51\fs18$,
$\rm{Dec.}=43\degr 54\arcmin 17\farcs85$.
The black contours are at n$\sigma$ ($n=3, 9, 27,81$ and $\sigma=0.2$~mJy~beam$^{-1}$).
The stars, circles, and a triangle are corresponding to the coordinates shown in Table \ref{coord}.
The stars represent the positions of the northern and southern nuclei.
The blue circle and triangle are the positions of the blue-shifted components, and the red circle is the positions of red-shifted components identified in section \ref{ID-OF-sec}.
We obtain the blue and red wing maps by integrating the emission in the velocity range [-360,-180] km s$^{-1}$ and [240,360] km s$^{-1}$.}

\label{Chan-HCN}
\end{center}
\end{figure*}

\begin{figure*}
\begin{center}
\includegraphics[width=17.5cm]{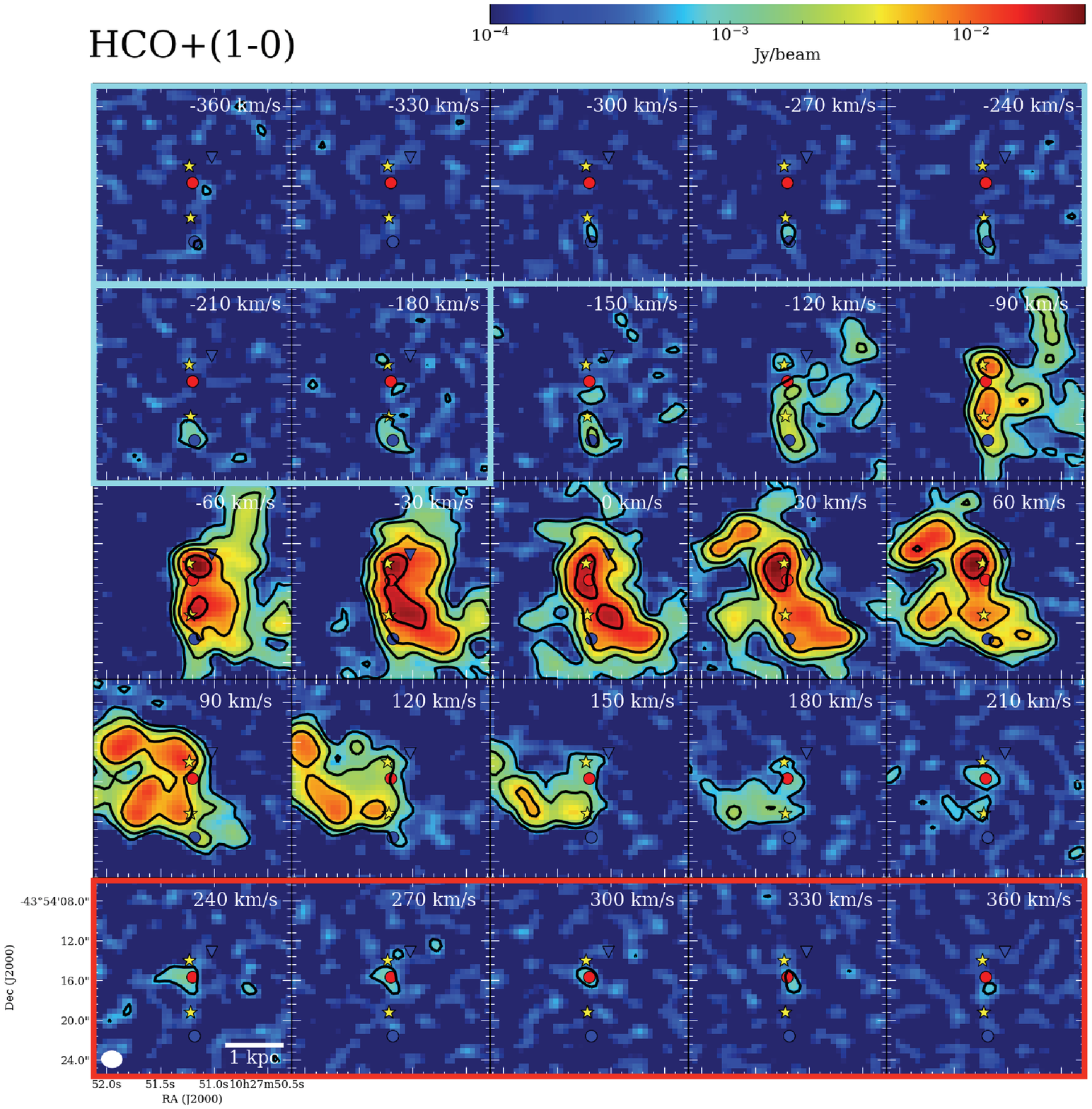}
\caption{
HCO$^+$~(1--0) channel map.
The size of each map is 20~\arcsec (3.4~kpc) square centered on
$\rm{RA.}=10^{\rm{h}} 27^{\rm{m}} 51\fs18$,
$\rm{Dec.}=43\degr 54\arcmin 17\farcs85$.
The black contours are at n$\sigma$($n=3, 9, 27,81$ and $\sigma=0.2$~mJy~beam$^{-1}$).
The stars, circles, and a triangle are corresponding to the coordinates shown in Table \ref{coord}.
The stars represent the positions of the northern and southern nuclei.
The blue circle and triangle are the positions of the blue-shifted components, and the red circle is the positions of red-shifted components identified in section \ref{ID-OF-sec}.
We obtain the blue and red wing maps by integrating the emission in the velocity range [-360,-180] km s$^{-1}$ and [240,360] km s$^{-1}$.}
\label{Chan-HCOp}
\end{center}
\end{figure*}

The HCN~(1--0) and HCO$^+$~(1--0) observations towards NGC~3256 were carried out on 2016, March 7 and 8 using 45 twelve meter antennas with the Band~3 receiver (ALMA Project ID: 2015.1.00993.S).
The range of unprojected baseline length is from 15 m to 460 m. The system temperature varied from 40 K to 100 K, and the total observation time is $\sim$ two hours including overhead for calibration (the on source time is $\sim$ 33 minutes).
We used the calibrated visibility data, and subtracted the continuum by using the uvcontsub task in ${\tt CASA}$ \citep{McMullin+07}.
We made data cubes with a velocity resolution of 30 km s$^{-1}$ using tclean task in ${\tt CASA}$ (version 4.7.0, Briggs weighting, robust = 0.5) with automatically masking loop as described in the ${\tt CASA~Guides}$\footnote{ \url{https://casaguides.nrao.edu/index.php/M100_Band3_Combine_4.3}}.
The r.m.s noise level is $\sim 0.2$ mJy~beam$^{-1}$ for the velocity resolution of 30~km s$^{-1}$ and the achieved beam size is $2\farcs1\times1\farcs7$ (360~$\times$~460~pc).
Channel maps are shown in Figures \ref{Chan-HCN} and \ref{Chan-HCOp}.
We define North-Nucleus and South-Nucleus (stars in Figures) as the peaks of the radio emission \citep{Neff+03}.
We made beam-matched ($2\farcs1\sim$ 360 pc) and uv-clipped ($>$ 18 k$\lambda$) images  so that their largest angular scales are consistent with those of the archival CO data.
The absolute flux density is calibrated using Ganymede, and the nominal calibration error is 5$\%$. The long term flux variation of the bandpass calibrator is 6$~\%$. 
Hence, we adopt 10$~\%$ as a conservative overall photometric error. 
In addition, we use the archival CO~(2--1) data that was obtained during Cycle 3 (ALMA Project ID: 2015.1.00714.S).
We made beam-matched and uv-clipped images by using clean task in ${\tt CASA}$ (version 4.7.0, Briggs weighting, robust = 0.5) with automatically masking loop.
Channel maps are shown in Figures \ref{CH-CO21}.
The CO~(1--0) and CO~(3--2) observations were conducted during Cycle 0 \citep{Sakamoto+14}(ALMA Project ID: 2011.0.00525.S).
We used the calibrated archival data and combined the visibility data from the compact (15--269 m) and the extended (16--374 m) configurations. 
These channel maps are shown in Figures \ref{CH-CO10} and \ref{CH-CO32}.

\begin{deluxetable*}{cccc}
\tablecaption{The coordinates of symbols in figures \label{coord}}
\tablewidth{0pt}
\tablehead{
region    &  RA.~(J2000) &  DEC.~(J2000) & symbol
}
\startdata
North-OF-Blue & $10^{\rm{h}}27^{\rm{m}}51\fs02$ & $-43\degr54\arcmin13\farcs12$ & blue triangle\\
North-Nucleus & $10^{\rm{h}}27^{\rm{m}}51\fs23$ & $-43\degr54\arcmin14\farcs00$ & star \\
OF-Red & $10^{\rm{h}}27^{\rm{m}}51\fs20$ & $-43\degr54\arcmin15\farcs64$ & red circle\\
South-Nucleus & $10^{\rm{h}}27^{\rm{m}}51\fs22$ & $-43\degr54\arcmin19\farcs20$ & star\\
South-OF-Blue & $10^{\rm{h}}27^{\rm{m}}51\fs17$ & $-43\degr54\arcmin21\farcs43$ & blue circle\\
\enddata
\end{deluxetable*}

\section{Identification and properties of molecular outflows}
\subsection{Outflow Identification}
\label{ID-OF-sec}
\begin{figure*}
\begin{center}
\includegraphics[width=17cm]{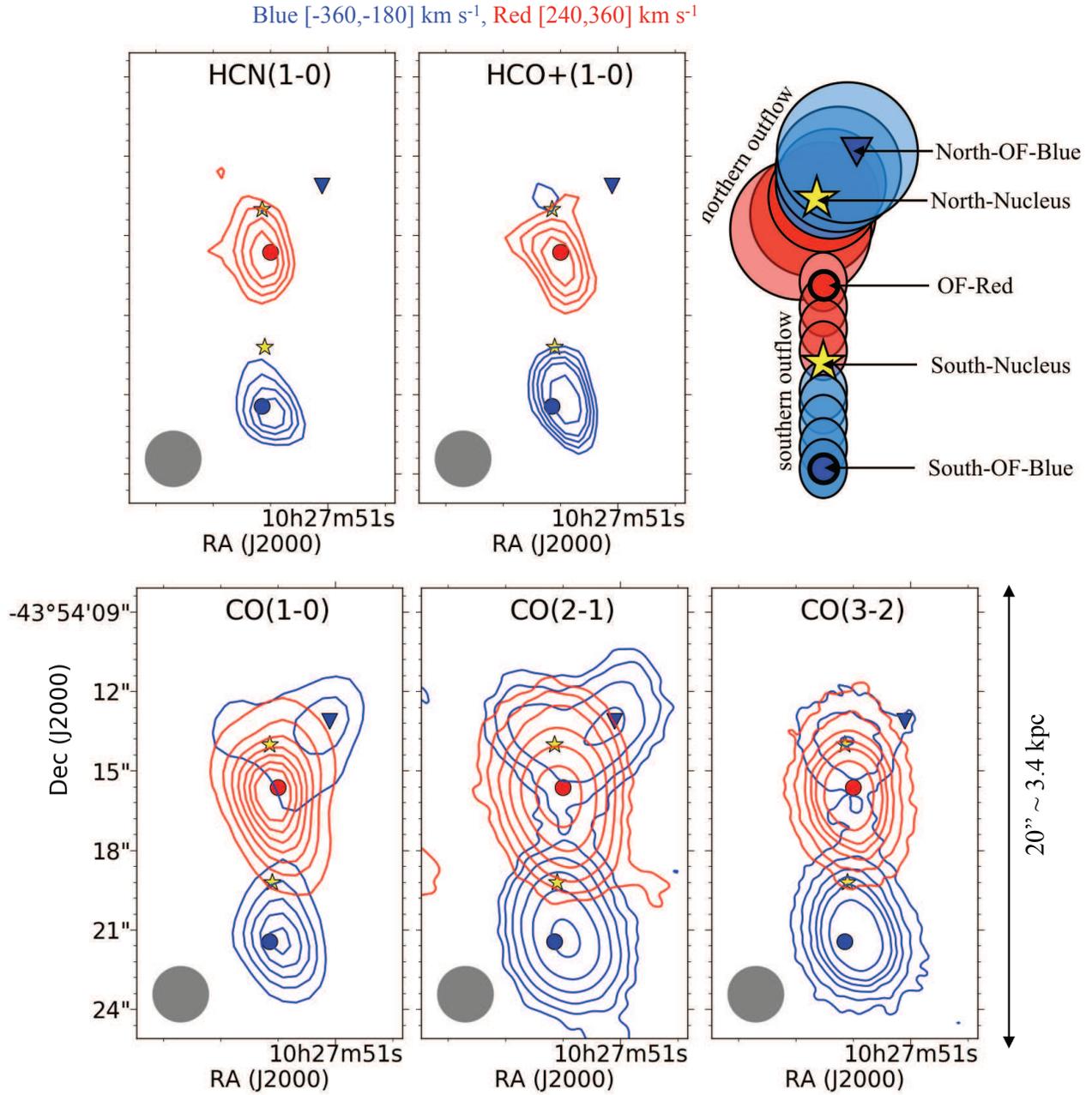}
\caption{The spatial distribution of the extreme velocity components.
The beams are restored by $2\farcs1$. 
The size of these images are 10 \arcsec $\times$ 20 \arcsec (1.7~$\times$~3.4~kpc) rectangle centered on
$\rm{RA.}=10^{\rm{h}} 27^{\rm{m}} 51\fs18$,
$\rm{Dec.}=43\degr 54\arcmin 17\farcs85$.
Blue- and red-shifted gas are shown by blue and red contour respectively.
The stars and circles correspond to the coordinates shown in Table \ref{coord}.  The contours are
$0.02\times(3,4,5,6)$ [Jy beam$^{-1}$ km s$^{-1}$] for HCN~(1--0), 
$0.02\times(3,4,5,6,8 )$ [Jy beam$^{-1}$ km s$^{-1}$] for HCO$^{+}$~(1--0),
$0.1\times(3,6,9,12,15,18,21)$ [Jy beam$^{-1}$ km s$^{-1}$] for CO~(1--0), 
$0.07\times(3,9,27,54,81)$ [Jy beam$^{-1}$ km s$^{-1}$] for CO~(2--1), and
$0.2\times(3,9,18,27,36,54)$ [Jy beam$^{-1}$ km s$^{-1}$] for CO~(3--2).
At top right panel, we show the schematic view of the outflows.
These outflows are predicted by \citet{Sakamoto+14} based on CO observations.
There are two bi-polar outflows (northern outflow and southern outflow), and we assume that red- and blue-shifted extreme velocity components of HCN~(1--0) and HCO$^+$~(1--0) is associated with southern outflow.
}
\label{OF-mom0}
\end{center}
\end{figure*}

The main goals of this paper are to investigate the properties of the molecular outflows using dense gas tracers and to derive the physical parameters from multi-line analysis.
As a first step, we use the HCN~(1-0) and HCO$^+$~(1--0) maps to identify the emission associated with outflow.
In general, the kinematics of the outflowing gas should be distinct from the systematic rotation of the galaxy.
Therefore, a natural simplification method to isolate the gas associated with the outflow is to extract the emission at the extreme (i.e. largest and smallest) velocities in the channel map.
We define the emission in
$v=[240,360]~\rm{km~s}^{-1}$
and
$v=[-360,-180]~\rm{km~s}^{-1}$
as the red-shifted and blue-shifted outflow gas, respectively.
The moment 0 maps generated using these velocity range are shown in Figure~\ref{OF-mom0}.
The flux of HCN~(1--0) is 0.16 and 0.12 Jy~km~s$^{-1}$ whereas HCO$^+$~(1--0) is 0.16 and 0.20 Jy~km~s$^{-1}$ for the red- and blue-shifted emission, respectively.
In addition, we integrate the same velocity range of the CO~(1--0), CO~(2--1), and CO~(3--2) emission line (Figure~\ref{OF-mom0} bottom).

We define the blue-shifted component which is located at south of South-Nucleus as ``South-OF-Blue", and the red-shifted component which is located between North- and South-Nucleus as ``OF-Red".
Furthermore, we define the peak of the blue-shifted CO~(1--0) emission located $\sim$2\farcs4 north-west of the North-Nucleus as ``North-OF-Blue".
The coordinates of North-OF-Blue, North- Nucleus, OF-Red, South-Nucleus and South-OF-Blue are shown in Table~\ref{coord}.
The three outflow components identified here support the existence of two bi-polar outflows \citep{Sakamoto+14}, and a schematic is shown in Figure~\ref{OF-mom0} (upper right).  
We use the term ``northern outflow" to refer to the outflows from the northern nucleus, and ``southern outflow" from the southern nucleus.

There are several key differences between the distribution of dense gas tracers (i.e. HCN~(1--0) and HCO$^+$~(1--0)) and the relatively diffuse gas traced in the CO lines.
The clear difference is that the dense gas tracers are only detected at OF-Red and South-OF-Blue, whereas they are undetected at North-OF-Blue to a sensitivity of 0.02 Jy beam$^{-1}$ km~s$^{-1}$ (Figure~\ref{OF-mom0}).
We find that the CO~(1--0) fluxes in Figure~\ref{OF-mom0} are 5.3, 1.9 and 2.2 Jy~km~s$^{-1}$ for red-shifted emission around OF-Red, blue-shifted emission around North-OF-Blue, and blue-shifted emission around South-OF-Blue, respectively.
\footnote{
We note that the CO~(1--0) fluxes in \citet{Sakamoto+14} are 
8.9, 3.2, and 1.2 Jy~km~s$^{-1}$
for red-shifted emission around OF-Red, blue-shifted emission around North-OF-Blue, and blue-shifted emission around South-OF-Blue, respectively.
These values are different from our measurements.
The difference between our CO~(1--0) image and that of \citet{Sakamoto+14} is the uv-coverage, imaging parameters, and integrated velocity range.
For OF-Red and North-OF-Blue, the fluxes measured in \citet{Sakamoto+14} are larger than our measurements.
Since the data set used in \citet{Sakamoto+14} has compact uv-coverage than ours, these differences are due to spatially extended northern outflow.
For South-OF-Blue, the flux measured in \citet{Sakamoto+14} is smaller than our measurements.
This is due to the integrated velocity range.
When we use the velocity range of [-360,-240] for the blue-shifted components, the CO~(1--0) flux is $\sim$~1.1~Jy~km~s$^{-1}$, and our measurement is consistent with \citet{Sakamoto+14}.}
The CO~(1--0) flux at OF-Red is likely intermixed with the outflow gas from both the northern and southern nuclei, and disentangling the exact fractional contribution is non-trivial.
For simplicity, we assume that the flux ratio between the two blue shifted components (at North-OF-Blue and South-OF-Blue) is the fractional contribution of the southern outflow at OF-Red, which means that the CO~(1--0) line flux associated with the southern outflow is $\sim5$~Jy~km~s$^{-1}$.

\subsection{Outflow Properties}
We derive the dense gas mass ($M_{\rm dense}$) and molecular gas mass ($M_{\rm mol}$) associated with the outflows, assuming the luminosity to mass conversion factor $\alpha_{\rm HCN}$ and $\alpha_{\rm CO}$~[$M_\odot$~(K~km~s$^{-1}$ $\rm {pc}^2)^{-1}$];
\begin{eqnarray}
  &&\frac{M_{\rm dense}}{[M_\odot]}=
  \alpha_{\rm HCN}
  ~\frac{L_{\rm HCN(1-0)}}{[\rm K~km~s^{-1}~{pc}^2]}\\
  &&\frac{M_{\rm mol}}{[M_\odot]}=
  \alpha_{\rm CO}
  ~\frac{L_{\rm CO(1-0)}}{[\rm K~km~s^{-1}~{pc}^2]}
\end{eqnarray}
where $L_{\rm HCN(1-0)}$ and $L_{\rm CO(1-0)}$ are the luminosities of the outflow derived from the line flux derived in \ref{ID-OF-sec}.
The systematic error is estimated using the range of commonly adopted conversion factors; $\alpha_{\rm HCN}=[0.24,10.0]$ and $\alpha_{\rm CO}=[0.8,4.6]$
\citep{Bolatto+13a,Barcos-Munoz+18}.
We estimate the age ($Age$), outflow rate ($\dot{M}$), and kinetic energy ($P_{\rm kin,OF}$) from the following equations.
\begin{eqnarray}
&& \frac{Age}{{\rm Myr}} = \left(\frac{l_{\rm OF}}{\rm pc}\right)\left(\frac{V_{\rm OF}}{\rm km~s^{-1}}\right)^{-1}\\
&& \frac{\dot{M}}{M_\odot\rm {~yr^{-1}}} = 10^6\left(\frac{M_{\rm dense}~{\rm or}~M_{\rm mol}}{M_\odot}\right)\left(\frac{Age}{\rm Myr}\right)^{-1}\\
&& \frac{P_{\rm kin,OF}}{{\rm erg~s^{-1}}} = (3\times10^{35})\left(\frac{\dot{M}}{{M_\odot~yr^{-1}}}\right)\left(\frac{V_{\rm OF}}{\rm km~s^{-1}}\right)^2
\end{eqnarray}
where the $V_{\rm OF}$ is the representative velocity of the southern outflow, $l_{\rm OF}$ is the distance from the nucleus to the outflow.
The representative velocity of the southern outflow is derived from the central value of the integrated range of blue-shifted components (Figure \ref{Chan-HCN}), which corresponds to the de-projected outflow velocity ($V_{\rm OF}$) of $\sim 1600$~km~s$^{-1}$
\footnote{We note that the maximum velocity of the southern outflow can be as long as $V_{\rm OF}\sim 2800~$km~s$^{-1}$ since the highest blue-shifted velocity is 480~km~s$^{-1}$ for CO~(2--1) (Figures \ref{CH-CO21}).} assuming an inclination of 80$^\circ$ \citep{Sakamoto+14}. 
The distances from the nucleus to the outflow ($l_{\rm OF}$) are measured using the coordinates shown in Table \ref{coord}.
The results are shown in Table \ref{properties}.
In addition, in order to investigate whether supernovae explosions can explain these outflow, we calculate the kinetic power injected by supernovae, by using the relationship derived by \citet{Veilleux+05};
\begin{eqnarray}
  \frac{P_{\rm kin,SF}}{{\rm erg~s^{-1}}} = 7\times10^{41}~\frac{SFR}{{M_\odot~yr^{-1}}}
\end{eqnarray}
We adopt SFR of $\sim6$ M$_{\odot}$~yr$^{-1}$ \footnote{We note that this value is higher than the SFR derived by infrared SED fitting ($\sim$2~M$_{\odot}$~yr$^{-1}$) conducted by \citet{Sakamoto+14} based on the method introduced in \citet{Murphy+11}.} for the $\sim3\farcs0$ region around the southern nucleus, which is derived from infrared SED fitting \citep{Lira+08}.

We find that $P_{\rm kin,OF}/P_{\rm kin,SF}$ is $>20\%$ and $>100\%$ using $M_{\rm dense}$ and $M_{\rm mol}$, respectively.
For both cases, the $P_{\rm kin,OF}/P_{\rm kin,SF}$ is larger than the nominal ratio for pure starburst driven outflow ($\sim1\%$: \citet{Cicone+14,Pereira-Santaella+18}),
suggesting that another source of energy such as an AGN is needed to account for the observed high value of $P_{\rm kin,OF}$.
This is consistent with the earlier claim based on CO outflow parameters and geometry by \citet{Sakamoto+14} that the southern outflow is most likely entrained by a radio jet from a weak or recently dimmed AGN in the southern nucleus.

\begin{deluxetable*}{lcc}
\tablecaption{Southern outflow properties derived from HCN~(1--0) and CO~(1--0) observations. \label{properties}}
\tablewidth{0pt}
\tablehead{
 &  \multicolumn{2}{c}{Southern Outflow} \\
Properties & HCN~(1--0) & CO~(1--0)
}
\startdata
$\alpha_{\rm CO}$ or $\alpha_{\rm HCN}$~[$\rm{M_\odot}$~(K km~s$^{-1}$ $\rm {pc}^2)^{-1}$]	&0.24--10	&0.8--2.1\\
$M_{\rm dense}$ or $M_{\rm mol}$~[$10^5~$M$_{\odot}$]	&2.8--120	&100--260\\
$V_{\rm OF}$~[km~s$^{-1}$]	&\multicolumn{2}{c}{1600}\\	
$l_{\rm OF}~[{\rm pc}]$	&\multicolumn{2}{c}{400}\\	
$Age$~[Myr]	&\multicolumn{2}{c}{0.25}\\	
$\dot{M}$~[$M_{\odot}$~yr$^{-1}$]	&1--48	&40--104\\
${\rm log}(P_{\rm kin,OF})$~[erg~s$^{-1}$]	&42.0--43.6	&43.5--43.9\\
\enddata
\end{deluxetable*}

\subsection{Line Ratio}
The line flux associated with outflow must be quantified in a consistent manner before the line ratios are evaluated.
For the two nuclei, we integrate the emission in all of the channels where the emission is detected with signal to noise ratio of $>3$, whereas we integrate only the extreme velocity channels defined in Section \ref{ID-OF-sec} (Figure~\ref{OF-mom0}) for the outflows.
For the latter, the velocity-integrated intensities for each line may not include the entire outflowing gas, since it excludes the channels closer to the systemic velocity which may contain the gas entrained in the outflow at lower velocity.
Nonetheless, the line ratios of two lines derived from the extreme velocity is a good representation of the physical condition in the gas associated with the outflow.
In the case of non-detection, we use the three sigma upper limit.

We show the results in Table \ref{FD}. Finally we calculate the line ratios, $R_{21/10}$, $R_{32/10}$, $R_{\rm HCN/CO}$, $R_{\rm HCO^+/CO}$, and $R_{\rm HCN/HCO^+}$ (Table \ref{Ratio}).
The $R_{\rm HCO^+/CO}$ (i.e. dense gas fraction) and the $R_{32/10}$ are both larger in the southern outflow (0.20$\pm$0.04 and 1.3$\pm$0.2, respectively) than in the southern nucleus (0.08$\pm$0.01, 0.7$\pm$0.1, respectively).
This trend is not seen in the case of the northern outflow.

\begin{deluxetable*}{ccccccc}
\tablecaption{Measured line flux in each region\label{FD}}
\tablewidth{0pt}
\tablehead{
region & velocity range & CO & CO & CO & HCN & HCO$^+$\\
 & & $J$=1--0  & $J$=2--1 &$J$=3--2 & $J$=1--0 & $J$=1--0\\
 & km s$^{-1}$ & \multicolumn{5}{c}{Jy km s$^{-1}$}
}
\startdata
North-OF-Blue &  [-360,-180] & 0.5 $\pm$ 0.1 & 1.0 $\pm$ 0.1 & $<$0.6  & $<$0.06  & $<$0.06  \\
North-Nucleus &  [-390,390] & 36 $\pm$ 4 & 139 $\pm$ 14 & 280 $\pm$ 28 & 2.6 $\pm$ 0.3 & 2.8 $\pm$ 0.3 \\
OF-Red & [240,360] & 1.5 $\pm$ 0.2 & 5.3 $\pm$ 0.5 & 9.5 $\pm$ 1 & 0.08 $\pm$ 0.02 & 0.08 $\pm$ 0.02 \\
South-Nucleus &  [-390,390] & 39 $\pm$ 4 & 142 $\pm$ 14 & 250 $\pm$ 25 & 1.1 $\pm$ 0.1 & 1.9 $\pm$ 0.2 \\
South-OF-Blue &  [-360,-180] & 0.8 $\pm$ 0.08 & 4.1 $\pm$ 0.4 & 9.9 $\pm$ 1.0 & 0.07 $\pm$ 0.02 & 0.10 $\pm$ 0.02 \\
\enddata
\tablecomments{The line fluxes are measured at inner $2\farcs1$ region in Figure~\ref{OF-mom0} for the extreme velocity components.}
\end{deluxetable*}

\begin{deluxetable*}{cccccc}
\tablecaption{Line ratios in each region \label{Ratio}}
\tablewidth{0pt}
\tablehead{
region  
  &$R_{21/10}$&
  $R_{32/10}$&
  $R_{\rm HCN/CO}$&
  $R_{\rm HCO^+/CO}$&
  $R_{\rm HCN/HCO^+}$
}
\startdata
\hline
North-Nucleus & 1.0$\pm$0.1 & 0.9$\pm$0.1 & 0.12$\pm$0.02 & 0.13$\pm$0.02 & 0.94$\pm$0.13 \\ 
North-OF-Blue & 0.5$\pm$0.1 & $<$0.1 & $<$0.20 & $<$0.20 & - \\ 
OF-Red & 0.9$\pm$0.1 & 0.7$\pm$0.1 & 0.09$\pm$0.02 & 0.09$\pm$0.02 & 0.99$\pm$0.34 \\ 
South-Nucleus & 0.9$\pm$0.1 & 0.7$\pm$0.1 & 0.05$\pm$0.01 & 0.08$\pm$0.01 & 0.60$\pm$0.08 \\ 
South-OF-Blue & 1.2$\pm$0.2 & 1.3$\pm$0.2 & 0.13$\pm$0.04 & 0.20$\pm$0.04 & 0.68$\pm$0.24 \\ 
\enddata
\end{deluxetable*}

\color{black}
\section{Radiative Transfer Modeling}
\label{RADEX-sec}
\subsection{${\tt RADEX}$ and Bayesian analysis}
The physical properties (e.g., temperature and density) of the outflow can be substantially different from the conditions seen in the ambient ISM. 
Thefore, a radiative transfer treatment is necessary to investigate the physical properties.
We use the non local thermodynamic equilibrium (non-LTE) radiative transfer code ${\tt RADEX}$ \citep{RADEX} to estimate the physical properties from the measured fluxes of CO~(1--0), CO~(2--1), CO~(3--2), HCN~(1--0), and HCO$^{+}$(1--0) for North-Nucleus, South-Nucleus, and South-OF-Blue.
We assume a background temperature of $T_{\rm bg}=2.73$ K.
We adopt a line width of 150~km~s$^{-1}$ (full width half maximum, FWHM) as a typical value since the FWHM of CO~(2--1) emission at North-Nuclei and South-Nuclei are 160 and 112 km s$^{-1}$, respectively.
We adopt the CO and H$_2$ abundance ratio of [CO/H$_2$] =$1.0\times10^{-4}$ assuming 15 $\%$ of C is in the form of CO with a cosmic value of (C/H)$ =3\times10^{-4}$ \citep{Omont+07}.
We assume the same FWHM = 150 km~s$^{-1}$ between the outflow and nuclei.
In order to check how the value of FWHM affects the results, we conducted the same calculation assuming 300~km~s$^{-1}$ and 50~km~s$^{-1}$.
We find that the results are consistent to within one sigma range.
Under these assumptions, we can estimate the expected line intensities from the kinetic temperature ($T_{\rm kin}$), H$_2$ volume density $(n_{\rm H_2}$), CO column density ($N_{\rm CO}$), beam filling factor (FF), and the abundance ratio ($N_{\rm HCN}$/$N_{\rm CO}$ and $N_{\rm HCO}^+$/$N_{\rm CO}$) using $\tt{RADEX}$.
We search the best parameters within a range of $T_{\rm kin}=5-6300$ K, $n_{\rm H_2}=10^1-10^7$ cm$^{-3}$, $N_{\rm CO}=10^{12}-10^{22}$ cm$^{-2}$, $FF=10^{-5}-10^{0}$, N$_{\rm HCN}$/N$_{\rm CO}$=$10^{-7}-10^{-3}$, and N$_{\rm HCO^+}$/N$_{\rm CO}$=$10^{-7}-10^{-3}$.
In order to constrain the free parameters ($T_{\rm kin}$, $n_{\rm H_2}$, $N_{\rm H_2}$, $FF$, $N_{\rm HCN}$/$N_{\rm CO}$ and $N_{\rm HCO}^+$/$N_{\rm CO}$), we conducted Bayesian likelihood statistical analysis, following the method described in \citet{Kame+12}.
We use the diameter of the photometric beam size ($\sim400$~pc$~\sim2.1\arcsec$) as the upper limit to the length of the column.
Similarly, we use the virial mass (${M_{\rm dyn}}=1040~R~{\it \sigma_v}^2$; ${\rm FWHM}= 2.35{\it \sigma_v}$, $R\sim$200~pc) \citep{Wilson+90} as the upper limit to the total mass.

\subsection {Results}
\begin{figure}
\begin{center}
\includegraphics[width=9cm]{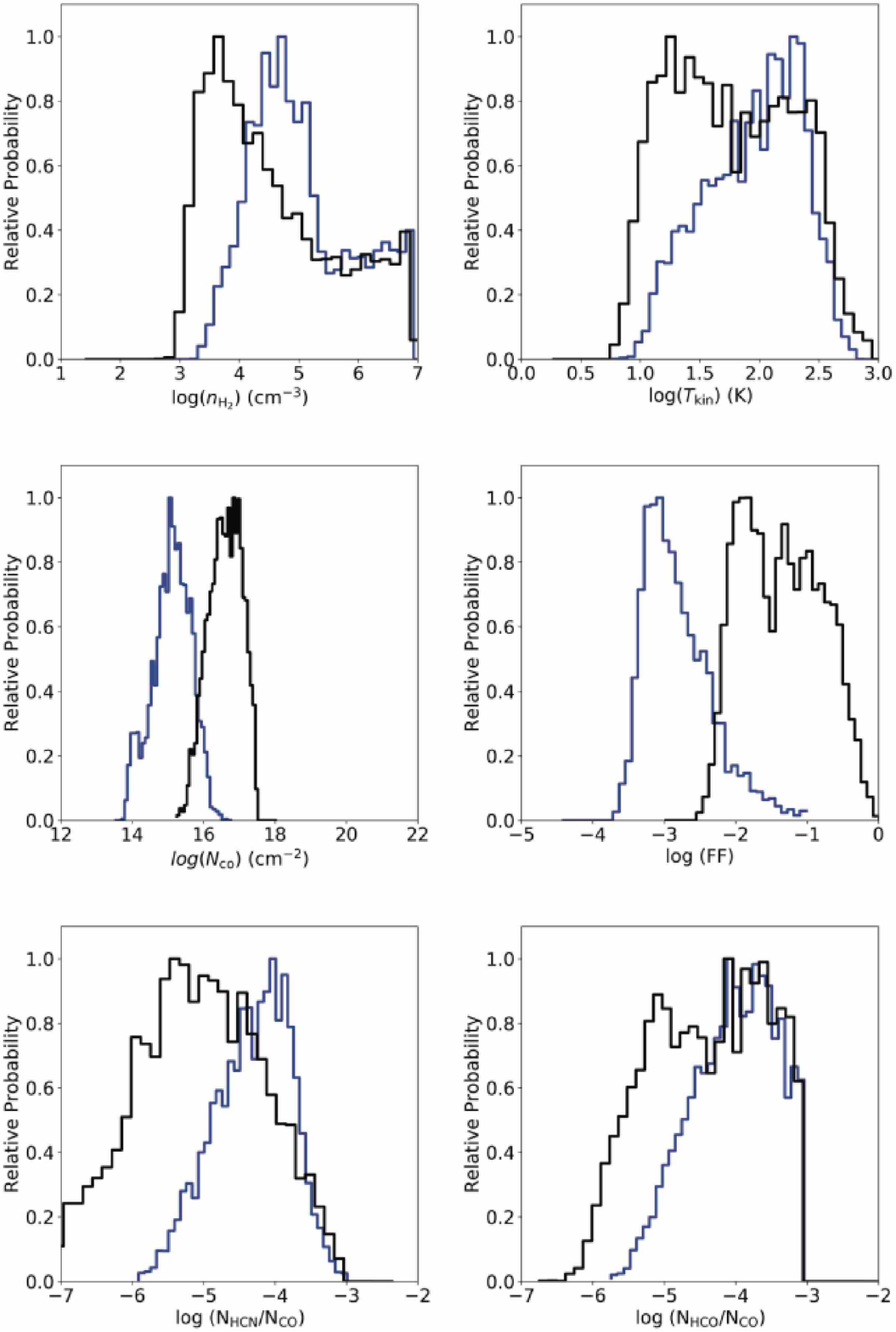}
\caption{The results of $\tt{RADEX}$ modeling. We show the probability functions for each parameter measured by CO, HCN, and HCO$^+$ flux.
The black lines show the relative probability function for each parameter at South-Nucleus.
The blue lines show the relative probability function for each parameter at South-OF-Blue.}
\label{RADEX-1}
\end{center}
\end{figure}

We show the results of the Bayesian likelihood statistical analysis in Table \ref{Bay}, and the relative probability function for the free parameters ($n_{\rm H_2}$, $T_{\rm kin}$, $N_{\rm CO}$, $FF$, $N_{\rm HCN}$/$N_{\rm CO}$ and $N_{\rm HCO}^+$/$N_{\rm CO}$) at South-Nucleus (black) and South-OF-Blue (blue) in Figure \ref{RADEX-1}.
Except for the $N_{\rm CO}$ and $FF$, the difference between South-OF-Blue and South-Nucleus are smaller than one sigma range \footnote{The one sigma range is estimated with the same manner conducted by \citet{Kame+12}}. (Table \ref{Bay}), and therefore the apparent offset in the peaks seen in Figure~\ref{RADEX-1} is mostly insignificant.
The critical issue of this modeling is that four lines are used as input parameters to estimate five parameters, resulting in the broad probability functions.
In addition, the assumption of a single phase ISM is not appropriate, if the ISM is composed of multiple phases (e.g., warm and cold).

\color{black}
\citet{Sliwa+14} and \citet{Saito+15} performed similar analyses using ${\tt RADEX}$ to investigate spatially resolved physical properties of the merging galaxies, NGC~1614 and VV~114.
\citet{Sliwa+14} show that the cold molecular gas component ($<$ 100 K) is dominant in NGC~1614.
In contrast, our estimated one sigma range of T$_{\rm kin}$ is warmer than 100 K.
While our one sigma range is larger than those derived by \citet{Sliwa+14} and \citet{Saito+15}, South-OF-blue shows warmer T$_{\rm kin}$ and smaller N$_{\rm{CO}}$ than every region in NGC~1614.
In order to estimate the physical properties at a higher accuracy, optically thin lines such as $^{13}$CO and/or high-$J$ lines are essential \citep{Sliwa+14,Saito+15}.
While there are some uncertainties, this analysis is an important first step to quantify the physical properties of extra-galactic molecular outflow.

\begin{deluxetable*}{ccccccc}
\tablecaption{The results of $\tt{RADEX}$ modeling \label{Bay}}
\tablewidth{0pt}
\tablehead{
region &  $T_{\rm kin}$ &  $n_{\rm H_2}$ & $N_{\rm CO}$ & FF & N$_{\rm HCN}$/N$_{\rm CO}$ & N$_{\rm HCO^+}$/N$_{\rm CO}$\\
        &  K &  cm$^{-3}$ & cm$^{-2}$ &  &
}
\startdata
South-Nucleus & 17 - 259 & $10^{3.6}$ - $10^{6.0}$ & $10^{18.4}$ - $10^{19.4}$ & 0.01 - 0.21 & $10^{-5.9}$ - $10^{-4.1}$ & $10^{-5.3}$ - $10^{-3.5}$ \\ 
South-OF-Blue & 33 - 254 & $10^{4.3}$ - $10^{6.2}$ & $10^{16.6}$ - $10^{17.7}$ & 0.0006 - 0.0055 & $10^{-4.9}$ - $10^{-3.8}$ & $10^{-4.7}$ - $10^{-3.3}$\\
\enddata
\tablecomments{The ranges are estimated by the one sigma range in the same manner as \citet{Kame+12}.}
\end{deluxetable*}

\section{Discussion}
\label{Discussion-sec}
The diffuse gas traced in the commonly observed CO line allows us to study the overall properties of the gas outflow such as the mass and kinematics.
As shown in this study, a multi-line analysis including the dense gas tracers allows us to derive important new information pertaining to the physical and chemical properties of the outflow.
Assuming that the HCN and HCO$^+$ are collisionally excited, the $R_{\rm HCN/CO}$
and
$R_{\rm HCO^+/CO}$
trace the dense gas fraction \citep[e.g.,][]{Gao+04,Lada+12,Bigiel+15, Maria+17}.
Although we interpret these ratios as dense gas fraction, we note that the HCN and HCO$^+$ emission can be enhanced due to turbulence \citep[e.g.,][]{Krumholz+05, Usero+15}.
In the following, we compare the ratios ($R_{\rm HCN/CO}$, $R_{\rm HCO^+/CO}$, $R_{\rm 21/10}$, and $R_{\rm 32/10}$) between the nucleus and outflow components for both the northern and southern galaxy, and investigate them in the context of the nuclear activity. 
We discuss each outflow component (i.e. South-OF-Blue, North-OF-Blue, OF-Red) separately in the following subsections.

\subsection{South-OF-Blue}\label{Discussion:SouthOutflow}

\begin{figure*}
\begin{center}
\includegraphics[width=17cm]{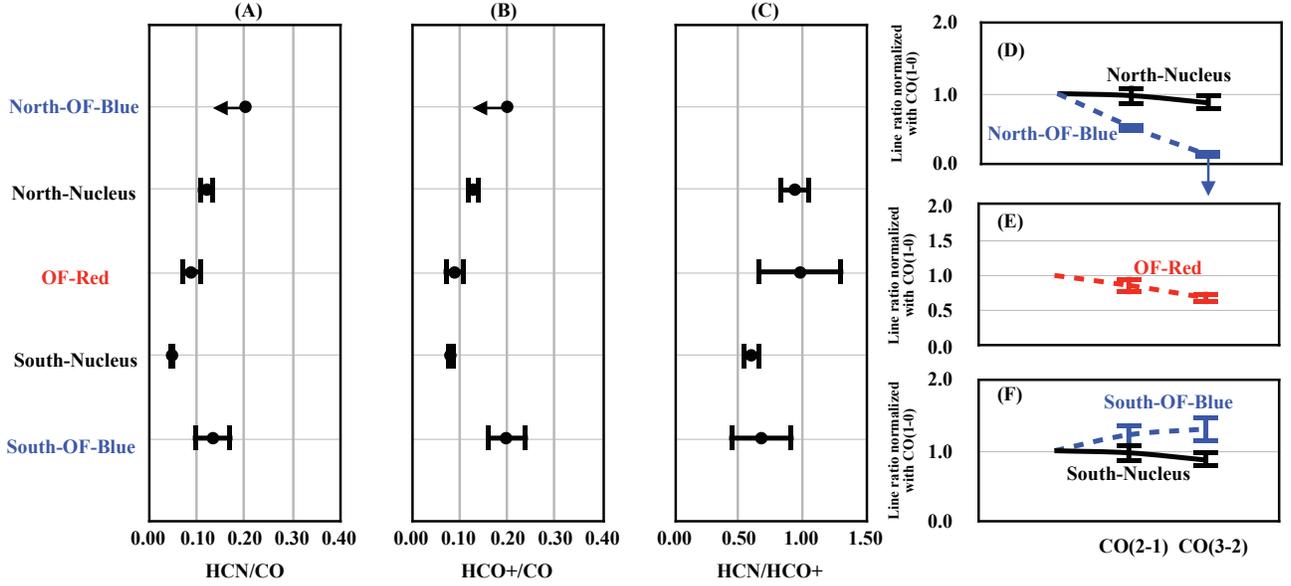}
\caption{(A)-(C) Line ratios for each region.
The panels show the
(A)~$R_{\rm HCN/CO}$, 
(B)~$R_{\rm HCO^+/CO}$, and 
(C)~$R_{\rm HCN/HCO^+}$. 
The dense gas fraction traced by $R_{\rm HCN/CO}$ and $R_{\rm HCO^+/CO}$ is similar or smaller at North-OF-blue and larger at South-OF-blue than the nucleus.
(D)-(F) CO spectral line energy distribution up to $J$=3 for each region;
(D) North-OF-Blue and North-Nucleus, 
(E) OF-Red, 
(F) South-OF-Blue and South-Nucleus. 
A decreasing trend is seen at North-OF-Blue, and an increasing trend is seen at South-OF-Blue.
The values are shown in Table \ref{Ratio}.
}
\label{RatioFig}
\end{center}
\end{figure*}

\begin{figure*}
\begin{center}
\includegraphics[width=8cm]{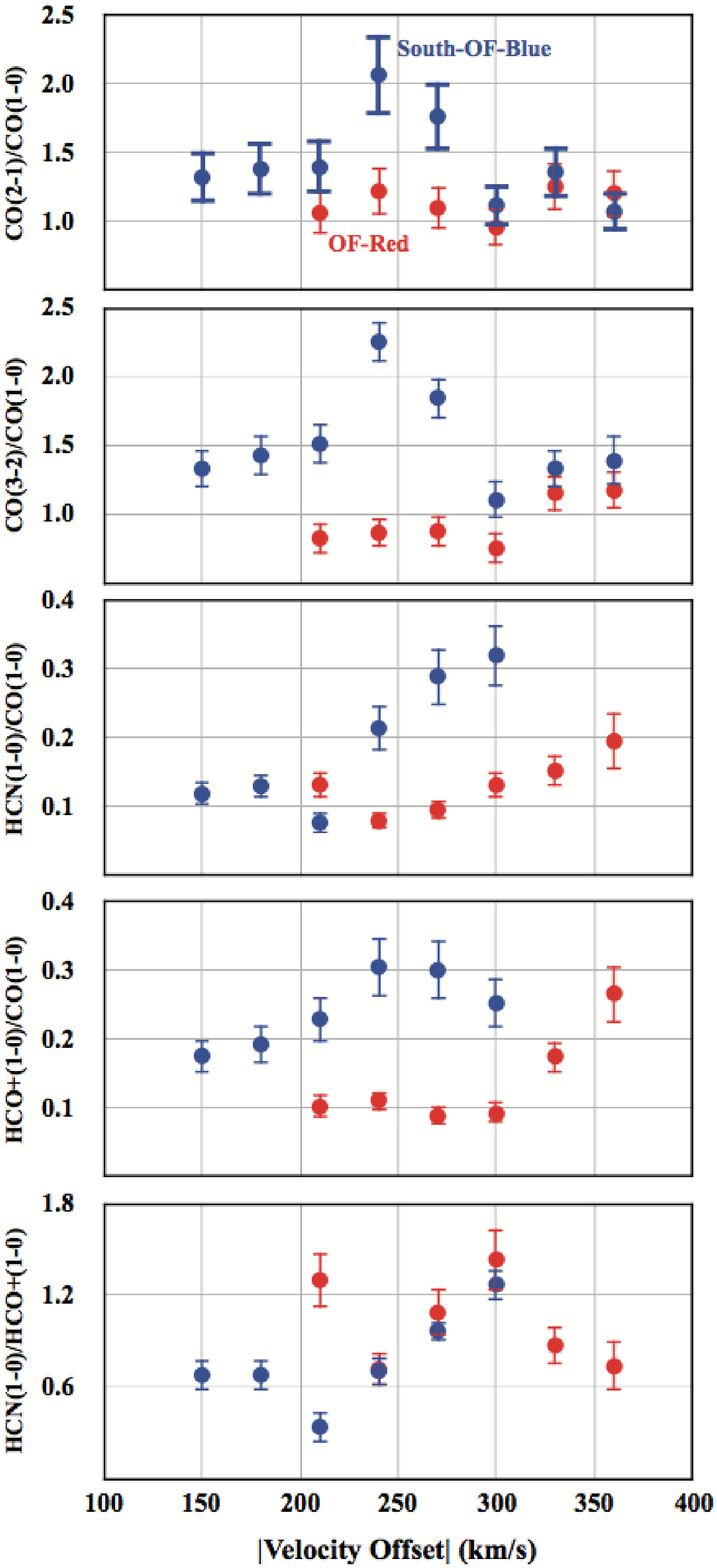}
\caption{
The relation between absolute value of velocity offset and the line ratios; 
(a)$R_{21/10}$,
(b)$R_{32/10}$,
(c)$R_{\rm HCN/CO}$,
(d)$R_{\rm HCO^+/CO}$, and 
(e)$R_{\rm HCN/HCO^+}$.
The blue shifted components (blue) are for South-OF-Blue, and red shifted components (red) are for OF-Red.
At South-OF-Blue, low-$J$ CO ratio decreases towards the largest velocity offset,
on the other hand, dense gas fraction increases.
This difference is probably due to two-phase ISM in the outflow.
In addition, $R_{\rm HCN/HCO^+}$ increases towards the largest velocity offset, and one possible explanation is that shock is dominant in the outflow.
}
\label{ratio-vel}
\end{center}
\end{figure*}

Dense gas traced in both HCN~(1--0) and HCO$^+$~(1--0) are clearly detected at South-OF-Blue (Figure~\ref{OF-mom0}).
We find that the dense gas fraction traced in $R_{\rm HCO^+/CO}$ at the South-OF-Blue ($R_{\rm HCO^+/CO}$ = $0.20\pm0.04$) is larger than the South-Nucleus ($R_{\rm HCO^+/CO}$ = $0.08\pm0.01$) by at least a factor of $\sim2$ (Table~\ref{Ratio} and Figure~\ref{RatioFig}).
The same trend is also seen in $R_{\rm HCN/CO}$.
In addition, we find that $R_{\rm 21/10} = 1.2\pm0.2$ and $R_{\rm 32/10}= 1.3\pm0.2$ at South-OF-Blue is larger than at South-Nucleus 
($R_{21/10}=0.9\pm0.1$ and $R_{32/10}=0.7\pm0.1$) by a factor of $\sim2$.
These results suggest larger dense gas fraction and higher excitation condition in the South-OF-Blue than in the South-Nucleus.
While the one sigma range is large, the higher $T_{\rm kin}$ and $n_{\rm H_2}$ (Figure~\ref{RADEX-1}) in South-OF-Blue compared to South-Nucleus further support the notion that the physical conditions of the ISM are significantly altered due to the energy input from the powerful outflow.

In order to further characterize the variation in the line ratio as a function of velocity, we derive the same line ratios using the channel map presented in Figure \ref{CH-Ratio-Blue} and \ref{CH-Ratio-Red}.
Figure \ref{ratio-vel} shows the line ratios plotted against the absolute value of velocity offset at OF-Red and South-OF-Blue.
At South-OF-Blue, we find that the dense gas fraction increases towards the largest velocity offset
(e.g.,
$R_{\rm HCN/CO}=0.22\pm0.03$ at -240 km~s$^{-1}$ 
and 
$0.32\pm0.05$ at -300 km~s$^{-1}$
)(Figure \ref{ratio-vel}).
In contrast, the $R_{32/10}$ and $R_{21/10}$ decrease towards the largest velocity offset
(e.g., $R_{32/10} = 2.3\pm0.2$ at -240 km~s$^{-1}$
and
$1.4\pm0.2$ at -360 km~s$^{-1}$
).

\begin{figure}
\begin{center}
\includegraphics[width=8cm]{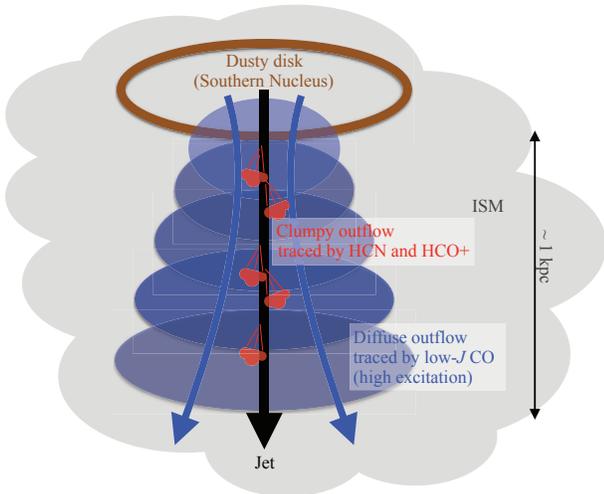}
\caption
{
A schematic view of expected feature of the blue-shifted components of southern outflow in NGC~3256 (South-OF-Blue).
There are clumpy and diffuse gas outflow.
The gas with the largest velocity offset is efficiently compressed into dense gas clumps due to a jet-ISM interaction.
Such clumps possibly lead future star formation.
}
\label{sketch}
\end{center}
\end{figure}

If HCN~(1--0), HCO$^+$(1--0), CO~(2--1) and CO~(3--2) trace the same gas, we can not explain such difference.
One possible explanation is a two-phase gas in the outflow; i.e., dense clumps are traced by HCN~(1--0) and HCO$^+$~(1--0) and diffuse gas are traced by CO lines.
We show the schematic view for the blue-shifted part of southern outflow in Figure \ref{sketch}.
In such a case, it is possible that the gas with the largest velocity offset is efficiently compressed into dense gas clumps while diffuse gas outflow is outspread.
This may explain the reason why the error bars in Bayessian estimation for the $T_{\rm kin}$ and $n_{\rm H_2}$ are large, as we assumed that the gas in the outflow consists of a single phase.

Numerical simulations suggest that multi-phase outflows occur due to the jet and ISM interaction (e.g., \citealt{Zubovas+14}; \citealt{Costa+15}; \citealt{Ferrara+16}).
This scenario is also supported by the detection of near-IR H$_2$ emission lines that trace gas heated by X-ray or shock in the southern outflow \citep{Emonts+14} and a collimated structure detected by the VLA \citep{Neff+03} whose spatial distribution is consistent with the outflow detected in CO \citep{Sakamoto+14} and dense gas tracers.

The characteristics of the southern outflow is similar to the strong radio jet seen in IC~5063 and M51.
IC~5063  \citep{Morganti+13,Tadhunter+14,Dasyra+15,Morganti+15,Dasyra+16} is a massive radio loud elliptical galaxy hosting a Seyfert 2 nucleus.
\citet{Tadhunter+14} suggest that the radio jet expansion drives the fast shock into the ISM from the observation of near-IR H$_2$ lines.
In addition, the
CO~(4--3)~/~CO~(2--1) in IC~5063
is over unity (in brightness temperature unit) at the gas associated with wind while the total
CO~(4--3)~/~CO~(2--1)
is $\sim 0.4$ \citep{Dasyra+16}.
This suggests high excitation condition in the jet driven outflows assuming optically thick gas.
In the case of M51, the outflow associated with the AGN jet shows large 
$R_{\rm HCN/CO}~>~1$ from a map obtained at $\sim34$~pc resolution at the Submillimeter Array (SMA) \citep{Matsushita+15}. 
They suggest that the HCN is enhanced in shocks.
If the dense gas is only associated with the shock front of the jet-ISM interaction in NGC~3256, the beam filling factor of HCN can be smaller than CO.
In such case, if the outflow is observed at 30~pc resolution in HCN, the dense gas fraction possibly increases ($>~1$) at the edge of the outflow.

The characteristics of the southern outflow are different from the outflow seen in Mrk~231, which has a molecular outflow associated with AGN.
In Mrk~231, broad wings have lower excitation condition than the core ($R_{21/10}\sim0.8$ in the core but $<$~0.8 in the outflow) \citep{Cicone+12}.
In order to explain the lower excitation conditions in the outflow in Mrk~231, \citet{Cicone+12} suggest acceleration by radiation pressure on the dust in molecular clouds \citep{Fabian+09} without the interaction of a radiation pressure driven jet with the ISM of the host galaxy.
In addition, $R_{\rm HCN/CO}$ in the outflow of Mrk~231 is higher ($\sim0.6$) \citep{Aalto+12} than the same quantity found in NGC~3256, although the reason is not clear.
The major difference between NGC~3256S and Mrk~231 is that NGC~3256S is not in a QSO phase as in Mrk~231, but it is a low luminosity AGN.

In summary our observations suggest that the southern molecular outflow has higher excitation condition and larger dense gas fraction than in the southern nucleus.
In addition, the southern molecular outflow is possibly represented by two-phases (dense clumps traced by HCN and HCO$^+$ and diffuse components traced by low-$J$ CO).
According to numerical simulations by \citet{Ferrara+16}, dense gas clumps represent only a transient phase ($<$~30~Myr) during the outflow evolution.
The next question is whether such dense clumps lead to star formation.
Numerical simulations show that jet-ISM interaction can lead to the formation of new massive stars \citep[e.g.,][]{Dugan+14, Wagner+16}. 
Assuming the relation between $SFR$ and dense gas mass
(SFR~[M$_\odot~\rm{yr}^{-1}]\sim10^{-7.93}\times M_{\rm dense}^{1.02}~[M_\odot]$),
we estimate $SFR$~=~0.003-0.14~[M$_\odot$~yr$^{-1}$] from dense gas mass associated with outflow ($M_{\rm dense}=(2.8-120)\times 10^5$~[M$_\odot$]).
A more direct observational evidence of positive feedback requires high resolution observations that trace the star formation.
Possible choices include optical and infrared emission lines \citep{Maiolino+17} and the hydrogen recombination lines seen in sub/millimeter wavelength (e.g., H42$\alpha$).

\subsection{North-OF-Blue}
Dense gas outflow traced in HCN~(1--0) and HCO$^+$~(1--0) is not detected at North-OF-Blue.
Our observation sensitivity is insufficient to derive a meaningful upper limit, making it difficult to compare the dense gas fraction between the northern nucleus and the outflow.
$R_{32/10}$ and 
$R_{21/10}$ are lower 
($R_{21/10}=0.5\pm0.1$, 
$R_{32/10}<0.1$)
at North-OF-Blue (dotted blue line in Figure \ref{RatioFig}~(D)) 
than the North-Nucleus
($R_{21/10}=1.0\pm0.1$, 
$R_{32/10}=0.9\pm0.1$) 
(solid line in Figure \ref{RatioFig} ~(D)).
Assuming that all CO lines are optically thick at North-Nucleus,
a possible explanation is that the outflow has lower or similar excitation condition compared to the northern nucleus.
This trend is similar to the starburst galaxy M82 that has
$R_{32/10}=1.1\pm0.2$
around the nucleus and $0.4\pm0.2$ in the outflow \citep{Weiss+05}.
Both the lower excitation and the smaller or similar dense gas fraction seen at the northern outflow in NGC~3256 suggest that the gas is expelled from the nucleus with little interaction between the outflow and the ISM.

Alternatively, the V-like shape in the moment 0 map of the blue-shifted CO~(2--1) component around North-Nucleus and North-OF-Blue (Figure \ref{OF-mom0}) suggests that it may arise from two different outflows, since the directions of the CO~(1--0) and CO~(3--2) outflow are different and faint CO~(3--2) is detected around North-OF-Blue (Figure \ref{OF-mom0}).
Such V-like shape outflow is reported at different scales: e.g., AGN jet in Circinus \citep{Marconi+94} and outflow cavity from low mass protostars \citep{Zhang+16}.
Further high resolution and sensitivity CO observations are necessary to test this idea.

\subsection{OF-Red}
Dense gas outflow traced in HCN~(1--0) and HCO$^+$~(1--0) is detected at OF-Red (Figures \ref{Chan-HCN}-\ref{OF-mom0}).
At OF-Red, both the southern and northern outflow can contribute to the red-shifted extreme gas components. 
For CO observations, we assume 54$\%$ and 46$\%$ of the red-shifted components arise from the southern and northern nucleus, respectively \citep{Sakamoto+14}.
On the other hand, since the dense gas tracers are not detected at North-OF-Blue, we assume that 100$\%$ of the red-shifted components arise from the southern nucleus in the case of HCN and HCO$^+$.
This indicates that the outflow emanating from the southern nucleus is denser than the northern outflow (Figure \ref{OF-mom0}).
We find that the dense gas fraction at OF-Red is a factor of a few smaller than at a South-OF-Blue at given absolute velocity offset (Figure \ref{ratio-vel}).
If we assume 54$\%$ of the CO red-shifted components arise from the southern outflow, then the dense gas fraction increases by a factor of two.
In such case, a symmetric bi-polar southern outflow between red- and blue-shifted components can be explained in terms of the dense gas fraction.
Furthermore, dense gas fractions increase at the largest velocity offset (Figure \ref{ratio-vel}).
For example, 
$R_{\rm HCN/CO}$ is 
$0.08\pm0.01$ at +240 km~s$^{-1}$ and 
$0.20\pm0.04$ at +360 km~s$^{-1}$.
This trend is also seen at South-OF-Blue and suggests that the gas is efficiently compressed into dense clumps with the largest velocity offset.

\subsection{Shock in the outflow}
We showed that the HCN and HCO$^+$ outflow detected in South-OF-Blue can be tracing dense clumps that are formed due to the jet-ISM interaction.
This possibly suggests the existence of shocks \citep{Goldsmith+17}, as suggested by \citet{Emonts+14} who show the possibility for the shock heated outflow from their near-IR H$_2$ observation. 
We investigate the difference between HCN~(1--0) and HCO$^+$~(1--0) outflows in terms of chemistry in the outflow in this section.

Shocks reduce the abundance of HCO$^{+}$ while leaving the HCN abundance unperturbed \citep{Iglesias+78,Elitzur+83,Mitchell+83,Harada+15}.
Strong UV radiation produced at the shock front can photo-dissociate CO, and make C available for further reaction networks leading to HCN formation \citep{Neufeld+89}. 
In the case of protostellar outflow, \citet{Burkhardt+16} show that the HCN and shocked gas tracers (e.g., CH$_3$OH, HNCO) can be seen at the edge of the outflow but HCO$^+$ is not.
\citet{Papa+07} suggest that the HCO$^+$ abundance can be reduced in the star forming and highly turbulent molecular gas found in LIRGs (Shocks are expected to be frequent in the highly supersonic turbulent molecular gas.).
In addition, PdBI observations towards Mrk~231 suggest that the HCN/HCO$^+$ ratio is enhanced in the outflow \citep{Lindberg+16}.
In the case of NGC~3256, $R_{\rm HCN/HCO^+}$
increases towards the largest velocity offset, i.e., $R_{\rm HCN/HCO^+}$ is 
$0.34\pm0.06$ at -240 km~s$^{-1}$ and 
$1.27\pm0.20$ at -360 km~s$^{-1}$ at South-OF-Blue. 
On the other hand, there is no clear enhancement of 
$R_{\rm HCN/HCO^+}$
towards the largest velocity offset at OF-Red in contrast to South-OF-Blue. 
This is possibly due to the orientation of the two galaxies,
i.e., a coexistence of the ISM belonging to the NGC~3256N and NGC~3256S toward the direction of OF-Red.
This means that enhancements of $R_{\rm HCN/HCO^+}$  can be seen over the entire extreme velocity components (from low to high velocity) at OF-Red.
We note that the red-shifted HNC wing is also detected at OF-Red \citep{Harada+18}, and HNC emission is expected in shock and/or dense warm gas \citep[e.g.,][]{Aalto+02}.

It is difficult to explain the origin of the large HCN/HCO$^+$ ratio seen in the outflows by shocks alone.
For example, \citet{Ziurys+89} and \citet{Mitchell+83} predict an increase in the HCO$^{+}$ abundance behind a shock front in a diffuse cloud ($\sim 3\times10^4$ cm$^{-3}$) through reaction of C$^{+}$ and O. 
The HCN and HCO$^+$ line emission alone are insufficient to draw a conclusion and observations of shocked gas tracers (e.g., CH$_3$OH, SiO, HNCO) in the outflow offer additional clues.
We simultaneously observed these molecules in the same data set presented here.
The flux is more than 10 times lower than HCN~(1--0) and HCO$^{+}$~(1--0), and the sensitivity was not enough to detect the extreme velocity components of these lines.
\citet{Harada+18} suggest the shock enhancement in the outflow based on the spatial distribution of these shock gas tracers.
Another way to identify the enhancement of shocks is to search for H$_2$O line in the outflow since H$_2$O is the most abundant molecule in the grain mantle \citep{Yamamoto+85} and H$_2$O abundance is enhanced if the grain mantels are progressively eroded by shock \citep{Flower+10}.
In addition, $Hershel$ observations show that the brightness of H$_2$O is comparable to high-$J$ CO \citep{Yang+13,Falstad+15}.
Recent ALMA detections of H$_2$O lines \citep{Konig+17,Pereira-Santaella+17} in LIRGs are promising, and future high resolution and sensitivity H$_2$O observations are the keys to investigate the shocks in extragalactic outflows.

\section{Summary}
\label{Summary-sec}
We present ALMA Cycle 3 observations of HCN~(1--0), HCO$^+$(1--0), and CO~(2--1) in NGC~3256.
We have detected the extreme velocity components showing outflow in all lines.
By comparing these data with the CO~(1--0) and CO~(3--2) ALMA Cycle 0 data, we measure line fluxes for
the northern nucleus (North-Nucleus),
the southern nucleus (South-Nucleus),
the northern blue-shifted gas (North-OF-Blue),
the southern blue-shifted gas (South-OF-Blue),
and red-shifted gas (OF-Red).
We investigate the line ratios ($R_{\rm HCN/CO}$, $R_{\rm HCO^+/CO}$, $R_{\rm 21/10}$, and $R_{\rm 32/10}$) for each position.
In addition we conduct $\tt{RADEX}$ modeling for South-Nucleus and South-OF-Blue.
Our findings are as follow,
\begin{itemize}
\item[(1)]
For the northern outflow, which is presumably triggered by starbursts,
HCN~(1--0) and HCO$^+$~(1--0) are not detected at North-OF-Blue.
$R_{\rm 21/10}$ and $R_{\rm 32/10}$ ratio are comparable between the outflow and the nucleus. 
One possible explanation is that the gas is expelled from the starburst region without experiencing large physical or chemical changes.
\item[(2)]
For the southern outflow emanating from the low-luminosity AGN in the southern nucleus, HCN~(1--0) and HCO$^+$~(1--0) are detected. 
Both the dense gas fraction and $R_{\rm 32/10}$ ratios are larger in the outflow than in the nucleus.
This suggests that the molecular outflow is warm, and gas becomes denser in the outflow.
\item[(3)]
By investigating line ratios for each velocity component in the southern blue-shifted outflow (at South-OF-Blue), we find that the dense gas fraction increases and $R_{\rm 32/10}$ decreases towards the largest velocity offset.
This suggests the existence of a two-phase (diffuse and clumpy) outflow.
One possible scenario to produce such a two-phase outflow is a jet-ISM interaction, which may trigger the shock and/or star formation in the outflow. 
\item[(4)] 
Shock is possibly enhanced in the outflow due to a jet-ISM interaction.
However, we need additional multi-molecule detections in the outflow to investigate chemical properties inside the outflow.
\end{itemize}

\acknowledgments
The authors thank the anonymous referee for carefully reviewing the manuscript.
We thank S.Aalto and E. Schinnerer.
They hosted TM for one month in Onsala observatory, Sweden and Max Planck Institute for Astronomy, Germany.
Authors thank S.Veilleux and R.Levy for useful discussion at the University of Maryland.
This work was supported in part by the Center for the Promotion of Integrated Sciences (CPIS) of SOKENDAI.
This paper makes use of the following ALMA data: ADS/JAO.ALMA $\#$2011.0.00525.S, 2015.1.00993.S, and 2015.1.00714.S.
ALMA is a partnership of ESO (representing its member states), NSF (USA) and NINS (Japan), together with NRC (Canada), MOST and ASIAA (Taiwan), and KASI (Republic of Korea), in cooperation with the Republic of Chile. The Joint ALMA Observatory is operated by ESO, AUI/NRAO and NAOJ.
This research has made use of the NASA/IPAC Extragalactic Database (NED) which is operated by the Jet Propulsion Laboratory, California Institute of Technology, under contract with the National Aeronautics and Space Administration. This research has made use of NASA's Astrophysics Data System.

%






\appendix
\section{Appendix information}

\renewcommand{\thefigure}{A-\arabic{figure}}
\setcounter{figure}{0}

\begin{figure}
\begin{center}
\includegraphics[width=15cm]{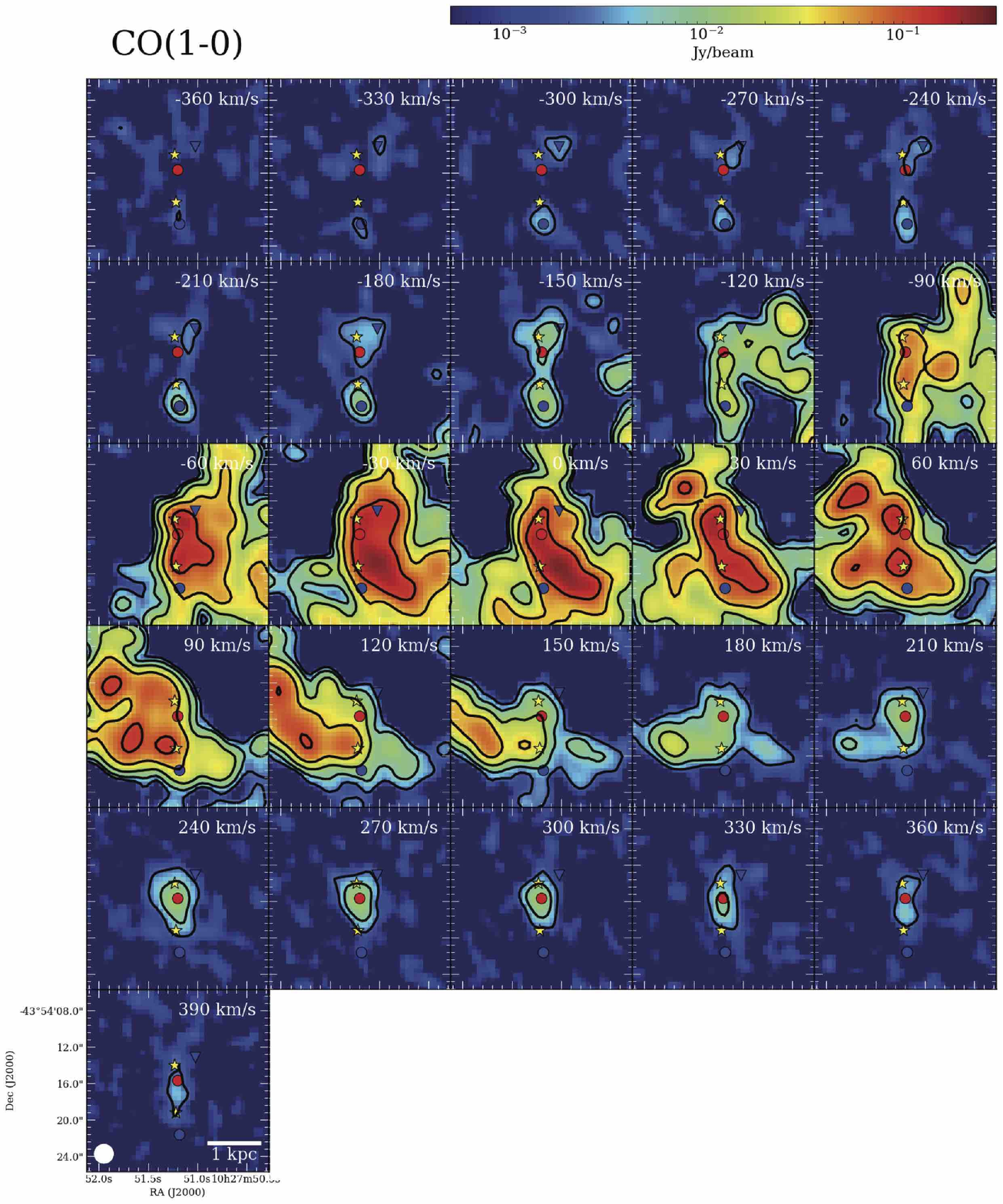}
\caption{CO~(1--0) channel map.
The size of each map is 20~\arcsec (3.4~kpc)} square centered on
$\rm{RA.}=10^{\rm{h}} 27^{\rm{m}} 51\fs18$,
$\rm{Dec.}=43\degr 54\arcmin 17\farcs85$.
The black contours are at n$\sigma$($n=3, 9, 27,81$ and $\sigma=3$~mJy~beam$^{-1}$).
The stars, circles, and a triangle are corresponding to the coordinates shown in Table \ref{coord}.
\label{CH-CO10}
\end{center}
\end{figure}

\begin{figure}
\begin{center}
\includegraphics[width=15cm]{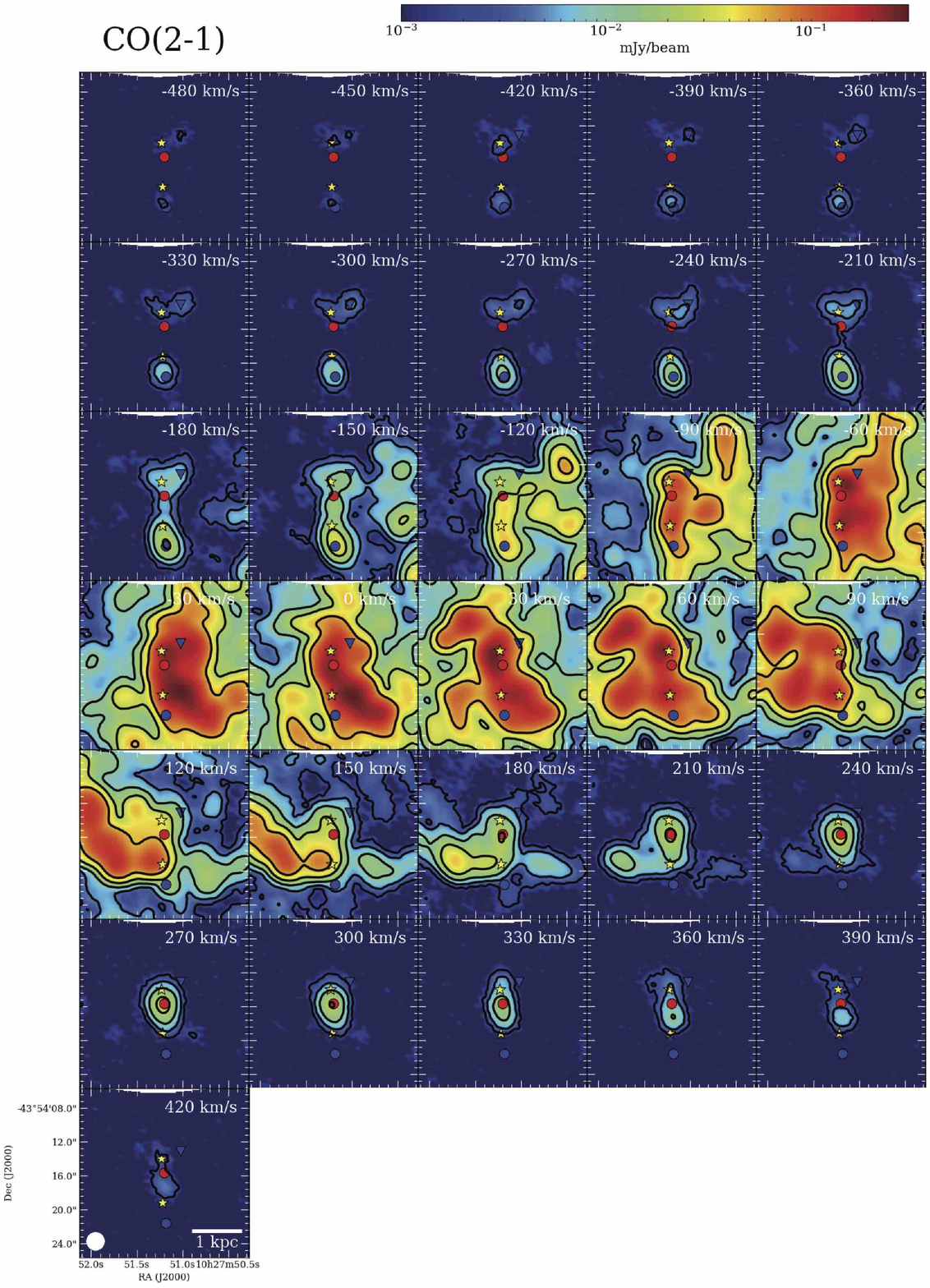}
\caption{CO~(2--1) channel map.
The size of each map is 20~\arcsec (3.4~kpc)} square centered on
$\rm{RA.}=10^{\rm{h}} 27^{\rm{m}} 51\fs18$,
$\rm{Dec.}=43\degr 54\arcmin 17\farcs85$.
The black contours are at n$\sigma$($n=3, 9, 27,81$ and $\sigma=2$~mJy~beam$^{-1}$).
The stars, circles, and a triangle are corresponding to the coordinates shown in Table \ref{coord}.
\label{CH-CO21}
\end{center}
\end{figure}

\begin{figure}
\begin{center}
\includegraphics[width=15cm]{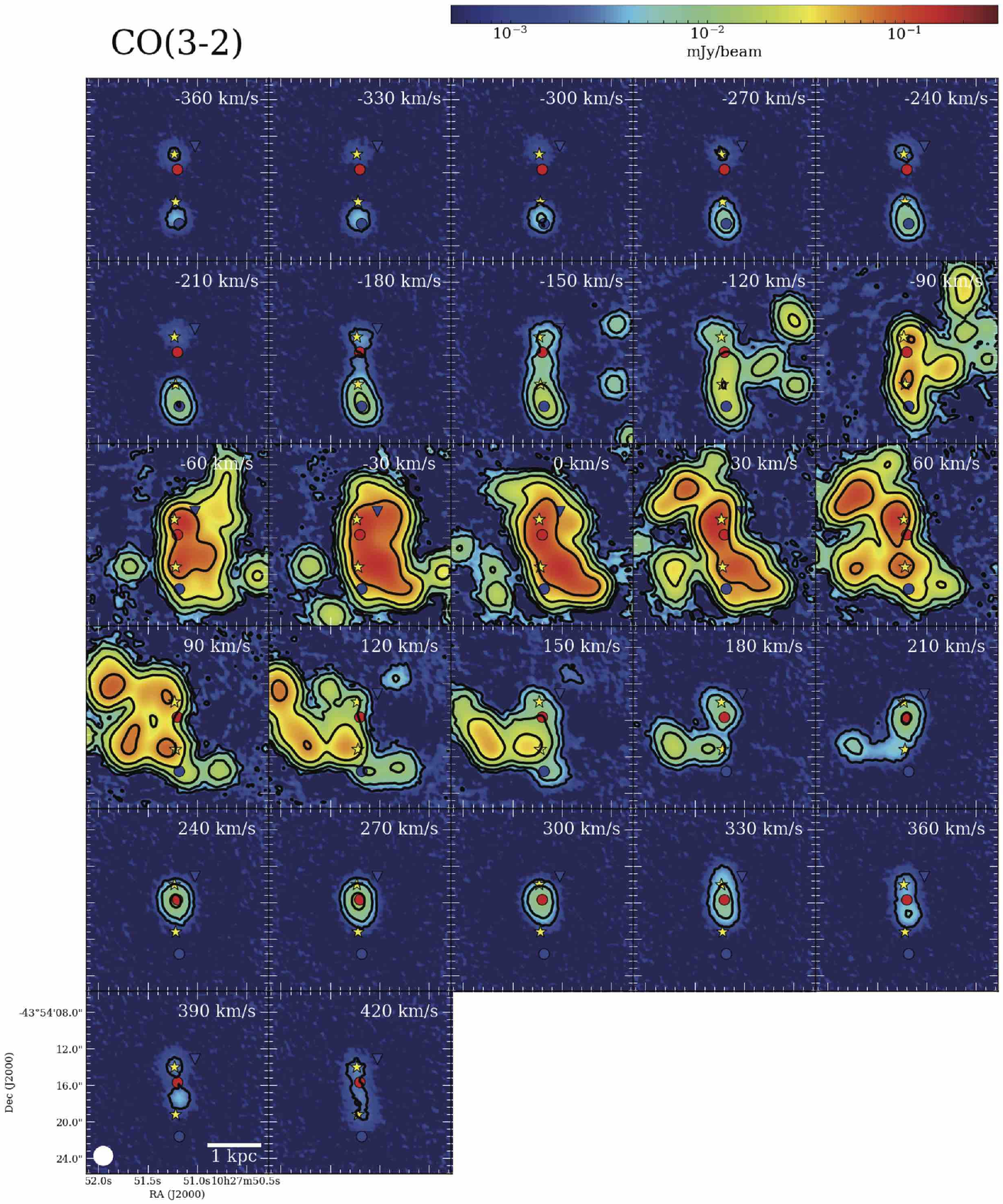}
\caption{
CO~(3--2) channel map.
The size of each map is 20~\arcsec (3.4~kpc)} square centered on
$\rm{RA.}=10^{\rm{h}} 27^{\rm{m}} 51\fs18$,
$\rm{Dec.}=43\degr 54\arcmin 17\farcs85$.
The black contours are at n$\sigma$($n=3, 9, 27,81$ and $\sigma=9$~mJy~beam$^{-1}$).
The stars, circles, and a triangle are corresponding to the coordinates shown in Table \ref{coord}.
\label{CH-CO32}
\end{center}
\end{figure}

\begin{figure}
\begin{center}
\includegraphics[height=21cm]{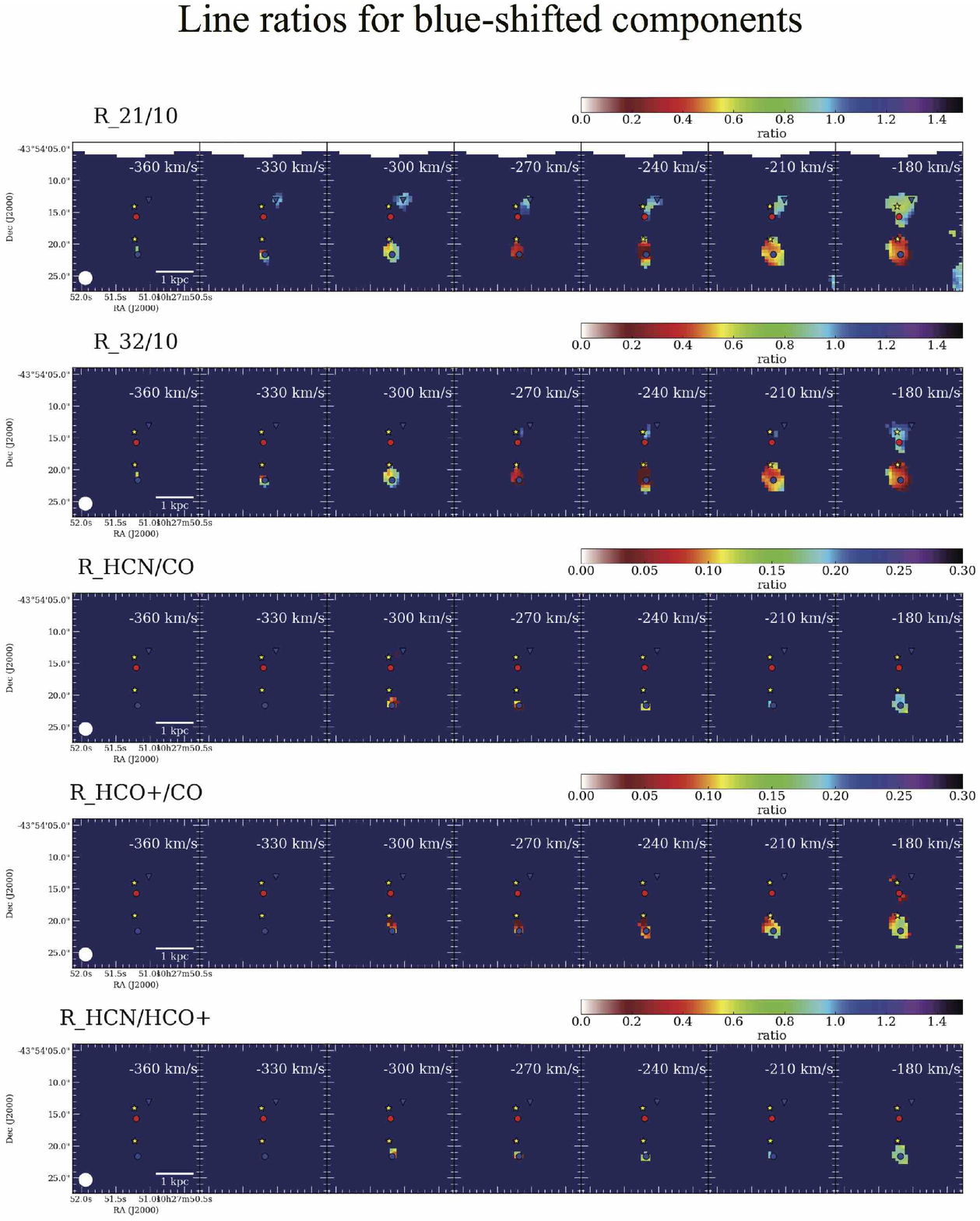}
\caption{Line ratios for blue-shifted components. 
Symbols are the same as Figure \ref{Chan-HCN}.
The size of each map is 20~\arcsec (3.4~kpc)} square centered on
$\rm{RA.}=10^{\rm{h}} 27^{\rm{m}} 51\fs18$,
$\rm{Dec.}=43\degr 54\arcmin 17\farcs85$.
    \label{CH-Ratio-Blue}
\end{center}
\end{figure}

\begin{figure}
\begin{center}
\includegraphics[height=21cm]{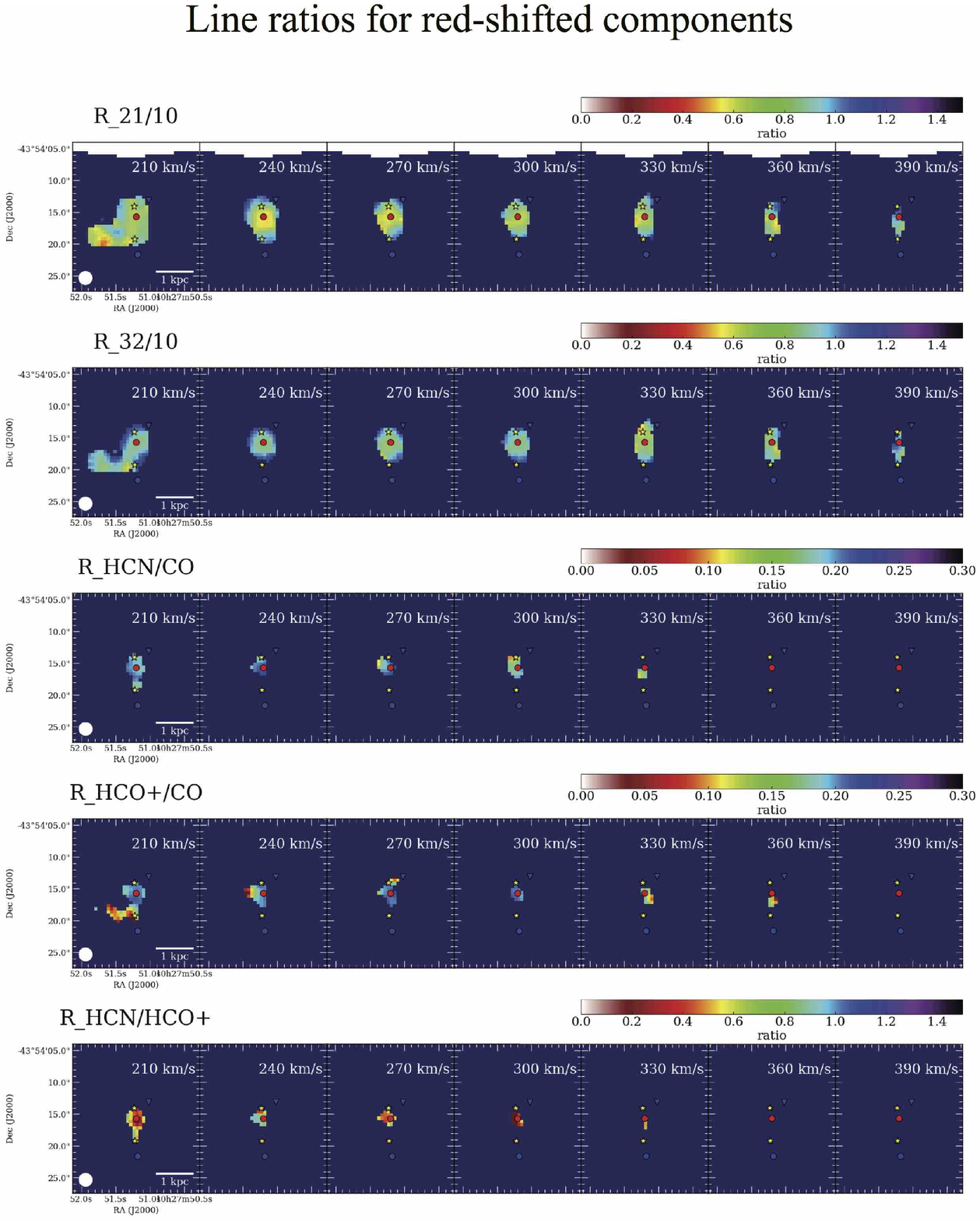}
\caption{Line ratios for red-shifted components. 
Symbols are the same as Figure \ref{Chan-HCN}.
The size of each map is 20~\arcsec (3.4~kpc)} square centered on
$\rm{RA.}=10^{\rm{h}} 27^{\rm{m}} 51\fs18$,
$\rm{Dec.}=43\degr 54\arcmin 17\farcs85$.
    \label{CH-Ratio-Red}
\end{center}
\end{figure}

\end{document}